\documentclass[conference]{IEEEtran}
\usepackage[top=0.70in, left=0.67in, bottom=1in, right=0.67in, columnsep=0.23 in]{geometry}

\usepackage[noend]{algpseudocode}

\usepackage{amssymb,algorithm,algpseudocode,amsmath,amsfonts,bm,epsfig,graphicx,theorem,epstopdf,wasysym,tabularx}
\usepackage{rotating,setspace,latexsym,epsf,color,dsfont,caption}
\usepackage{cite,subfigure}
\usepackage{algorithm,algpseudocode}

\allowdisplaybreaks

\newtheorem{Theorem}{Theorem}

\newtheorem{Lemma}{Lemma}
\newtheorem{Proposition}{Proposition}

\newenvironment{Proof}[1]{\medskip\par\noindent
{\bf Proof:\,}\,#1}{{\mbox{\,$\blacksquare$}\par}}

\newcommand{\vv}{\mathbf{v}}
\newcommand{\uv}{\mathbf{u}}
\newcommand{\yv}{\mathbf{y}}

\newcommand{\wv}{\mathbf{w}}

\newcommand{\Eb}{\mathbb{E}}
\newcommand{\Pb}{\mathbb{P}}

\title{Adaptive Coding for Information Freshness in a Two-user Broadcast Erasure Channel}
\author{
\IEEEauthorblockN{Songtao Feng \qquad Jing Yang}
\IEEEauthorblockA{School of Electrical Engineering and Computer Science \\
The Pennsylvania State University\\
University Park, PA 16802\\
\emph{\{sxf302,yangjing\}@psu.edu}}

\thanks{This work was supported in part by the US National Science Foundation (NSF) under Grant ECCS-1650299.}
}

\begin{document}
\IEEEoverridecommandlockouts
\maketitle
\thispagestyle{empty}

\begin{abstract}
In this paper, we investigate the impact of coding on the Age of Information (AoI) in a two-user broadcast symbol erasure channel with feedback. We assume each update consists of $K$ symbols and the source is able to broadcast one symbol in each time slot. Due to random channel noise, the intended symbol at each user will be erased according to an independent and identically distributed (i.i.d.) Bernoulli process. A user is able to successfully decode an update if it accumulates sufficient information and successfully decodes the $K$ symbols of the update. Assuming prefect feedback information at the source right after the transmission of each symbol, our objective is to design an adaptive coding scheme to achieve small AoI at both users. We propose a novel coding scheme to judiciously combine symbols from different updates together, and analyze the AoI at both users. Compared with a baseline greedy scheme, the proposed adaptive coding scheme improves the AoI at the weak user by orders of magnitude without compromising the AoI at the strong user.
\end{abstract}


\section{Introduction}\label{sec:intro}

In order to measure the ``freshness" of information, a metric {\it Age of Information} (AoI) has been introduced recently~\cite{infocom/KaulYG12}. A typical model to study AoI includes a source which generates time-stamped updates, and a destination which receives the updates transmitted over a network. The AoI of the destination, or simply the \emph{age}, is the time that has elapsed since the most recent update at the destination was generated at the source. More specifically, at time $t$, if the freshest update at the destination was generated at time $u(t)$, the age is $\delta(t):=t-u(t)$. 

Age of information as a ``freshness" metric has been studied in queueing systems with a single server~\cite{infocom/KaulYG12,ciss/KaulYG12,isit/YatesK12,YatesK16,Pappas:2015:ICC,isit/NajmN16,isit/KamKNWE16,isit/ChenH16}, and multiple servers~\cite{isit/KamKE13, isit/KamKE14,tit/KamKNE16}. For multi-hop networks, the optimality properties of a preemptive Last Generated First Served (LGFS) service discipline are established in~\cite{isit/BedewySS16}, and explicit age distributions based on a stochastic hybrid system approach are derived in~\cite{isit/Yates15}. AoI optimization has been studied in single-user systems~\cite{infocom/SunUYKS16,Improve,Wang:JCN:2019}, interfering links~\cite{He:TIT:2017}, multiple-access channels~\cite{Kaul:2017:MAC,Kadota:2018:INFOCOM} and broadcast channels~\cite{Modiano:2016:BC,Modiano:2018:BC}. AoI in energy harvesting systems has been analyzed in~\cite{ita/BacinogluCU15,Yang:2017:AoI,Arafa:TIT:2018,Feng:TAC:arxiv,Brown:2018:INFOCOM}.

Recently, coding for AoI optimization has received increasing attention. In~\cite{Yates:2017:ISIT}, two different coding strategies, i.e., rateless codes and maximum distance separable (MDS) codes, are studied for both single-user and multiple-user systems. They show that if the redundancy is carefully optimized in response to the channel erasure rate, the AoI performance of MDS coding can match that of rateless coding. In~\cite{Feng:ICC:2019}, the optimal transmission of rateless codes for AoI minimization in a single-user erasure channel has been characterized. In \cite{Najm:coding:2019}, it proves that when the source alphabet and channel input alphabet have the same size, a LCFS with no buffer policy is optimal. For an energy harvesting erasure channel, \cite{Baknina:2018:CISS} shows that rateless coding with save-and-transmit scheme outperforms MDS based schemes. For streaming source coding system, the optimal prefix-free lossless coding scheme that minimizes the average peak AoI is proposed in \cite{Emina:2018}. The effect of codeword length on the average AoI is analyzed in \cite{Parag:WCNC:2017,Sac:2018:SPAWC}. 
In \cite{Chen:CodeBC:2019}, the benefits of network coding in a two-user broadcast packet erasure channel with updates from two streams are studied. It shows that coded randomized policies outperform their uncoded counterparts in terms of age.

In this paper, we consider a two-user broadcast symbol erasure channel. The source continuously broadcasts encoded symbols of status updates to two users. Due to random erasures over the link from the source to each user, the users may not be able to decode the intended update at the same time. Assuming perfect feedback information at the source so it knows exactly which encoded symbols have been delivered to which user(s), our objective is to design an adaptive coding scheme to judiciously generate the encoded symbol each time, so that both users can successfully decode updates from the source timely. Intuitively, there exists a tension between the AoI of the two users. To see this, consider the scenario where the channel between the source and one user (termed as the strong user) is statistically better than the other (termed as the weak user). Assume the source prioritizes one of the users and will immediately switch to a new update once the previous update has been successfully decoded by the user with priority. Depending on whether the strong or weak user has the priority, there exist two different situations: If the strong user has the priority, then, with higher probability, the weak user will not be able to decode the same update when the strong user decodes. Thus, its AoI will keep growing until it eventually decodes one update successfully, leading to a larger AoI. On the other hand, if the weak user has the priority, then, with higher probability, the strong user will decode before the weak user decodes. The source will continue transmitting the same update until the weak one decodes. Compared with the first case, the AoI at the weak user will be lower, at the price of increasing the AoI at the strong user due to waiting. In analogy to the capacity region of broadcast channels, in this status updating setting, all achievable AoI pairs at both users form a region. While a complete characterization of such an \emph{AoI region} seems too ambitious at this stage, as a first step, we will investigate certain ``achievable points" within the AoI region under specific coding and transmission schemes.

Specifically, we will consider two updating schemes: a greedy scheme that always prioritizes the strong user, and another adaptive coding scheme that prioritizes the strong user but also takes the information freshness at the weak user into consideration. We aim to show that, compared with the greedy scheme, the adaptive coding scheme strictly improves the AoI at the weak user without compromising the AoI at the strong user. Such improvement becomes more prominent when the size of updates increases.

\section{System Model}
We consider a broadcast symbol erasure channel consisting of one source and two users. The source keeps generating updates of $K$ information symbols, and is able to broadcast one {\it encoded} symbol to the users in each time slot.  Assume the link between the source and user $i$, $i=1,2$ is noisy and each broadcast symbol can be erased over the link independently according to an i.i.d. Bernoulli process. 
Let $x_t$ be the broadcast symbol at time slot $t$, and $y_{1,t}$ and $y_{2,t}$ be the corresponding received symbol at user 1 and user 2, respectively. Then, $y_{i,t}$ equals $x_t$ with probability $p_i$ and equals $\varnothing$ if $x_t$ is erased. Without loss of generality, we assume $p_1> p_2$, i.e., user 1 is the strong user while user 2 is the weak user. We also assume perfect feedback information at the source right after each transmission. Therefore, the source knows which symbols have been received at each user at any time.

Denote the update generated at the source at time $t$ as $\uv_t:=\{u_t(k)\}_{k=1}^K$, where $u_t(k)$ is the $k$-th information symbol. Let $\uv^t:=\{\uv_\tau\}_{\tau=1}^t$, $\yv_i^t:=\{y_{i,\tau}\}_{\tau=1}^t$ for $i=1,2$. Then, in general,
$ x_t=f_t({\uv}^t,\yv_1^{t-1},\yv_2^{t-1} )$,
where $f_t$ is the encoding function at time $t$.

At the end of time slot $t$, user $i$ tries to decode an update based on ${\yv}_i^t$. 
An update is successfully decoded if the $K$ information symbols of the update are successfully decoded. 
If $\uv_\tau$ is decoded at time $t$ at user $i$, the instantaneous AoI at user $i$, denoted as $\delta_i(t)$, will reset to $t-\tau$. If multiple updates are decoded simultaneously, $\delta_i(t)$ will be reset to the smallest age of the decoded updates. 

Let
\begin{align}
\Delta_i:=\limsup_{T\rightarrow\infty}\Eb\left[\frac{1}{T}\int_{0}^{T}\delta_i(t)dt\right]
\end{align}
be the expected long-term average AoI at user $i$. Our ultimate goal is to characterize the maximum $(\Delta_1,\Delta_2)$ region over all possible coding schemes. While this is an extremely challenging problem in general, in this paper, we focus on specific coding schemes and characterize the corresponding achievable AoI pairs $(\Delta_1,\Delta_2)$. Specifically, we aim to show that by adaptively combining symbols from different updates into an encoded symbol, the AoI at the weak user can be significantly improved.

\section{Greedy Scheme}\label{sec:greedy}
In this section, we introduce a baseline greedy scheme. We assume that the source adopts the infinite incremental redundancy (IIR) strategy in \cite{Yates:2017:ISIT}. Under the IIR strategy, each $K$-symbol update is encoded by a rateless code, such as a Reed-Solomon or a Fountain code. The source keeps broadcasting the encoded symbols of an update to both users. 
Under the greedy scheme, the source prioritizes the strong user (i.e., user 1) and aims to minimize its time average AoI $\Delta_1$. Thus, as soon as user 1 successfully decodes an update, the source will switch to a new update and start broadcasting it. We have the following observations.

\begin{Theorem}\label{thm:greedy}
Under the greedy scheme, $\Delta_1=\frac{K}{p_1}\left(\frac{3}{2}+\frac{1-p_1}{K}\right)$,  $\Delta_2=\Omega(K(1/q)^{K})$, where $q:=\frac{p_1p_2}{\left(1-\sqrt{(1-p_1)(1-p_2)}\right)^2}< 1$.
\end{Theorem}

The proof of Theorem~\ref{thm:greedy} is omitted due to space limitation. 

\section{Adaptive Coding Scheme}\label{sec:coding}
Next, we present a novel adaptive coding scheme to strictly improve the greedy scheme. Our intuition is that by adaptively combining information symbols from different updates, the AoI of user 1 won't be affected while the AoI at user 2 will be significantly reduced.

The adaptive coding and updating scheme works in cycles, where each cycle begins with a phase 1, possibly followed by a phase 2.
We use $\wv:=\{w_{1},w_{2}\}$ to indicate be the updates that the users intend to decode at current time slot. If $w_{1}=w_{2}$, they aim to decode the same update, and the system works in phase 1; otherwise, the system operates in phase 2. Initially, we set $\wv=(\uv_1,\uv_1)$ at $t=1$. We also use $k_i,i=1,2$ to track the total number of random linear equations involving $w_i$ that have been received by user $i$. let $\vv_t:=(v_{1,t},v_{2,t})$ be the transmission status in time slot $t$. If $v_{i,t}=1$, it indicates the symbol broadcast at time $t$ has been successfully received at user $i$; otherwise, it is erased. At the end of each time slot, the encoder will update $\wv$ and $k_1$, $k_2$ based on the received feedback $\vv_t$, and decide the coding strategy for $t+1$.

The coding scheme is elaborated as follows.
\begin{itemize}
    \item \textbf{Phase $1$:} In Phase $1$, the source adopts rateless codes to encode $w_1$ and transmits encoded symbols continuously until user $1$ receives $K$ encoded symbols and successfully decodes $w_{1}$ at the end of a time slot $t$. Then, $w_{1}$ will be reset to $\uv_{t+1}$. Depending on whether user 2 has decoded $w_2$ at the end of $t$ or not, there are two different scenarios.
\begin{itemize}
    \item[(a)] User 2 has decoded $w_2$ at time $t$. Then, $w_2$ will be reset to $\uv_{t+1}$. The system enters phase 1 of the next coding cycle at $t+1$.

\item[(b)] User 2 has not decoded $w_2$ yet. The system then enters phase $2$ at $t+1$. 
\end{itemize}
We note that during phase 1, $k_i$ will keep increasing according to $\vv_{i,t}$ until it reaches $K$; it will then be reset to zero if $w_i$ is changed to a new update.

\item \textbf{Phase $2$:} During phase 2, the source will broadcast two different types of symbols: 1) the $(k_1+1)$th uncoded information symbol of $w_1$, denoted as $w_1(k_1+1)$, or 2) a random linear combination of $w_1(k_1+1)$ and the symbols of $w_2$. Denote $c_t\in\{1,2\}$ as the type of symbols broadcast at time $t$. At the beginning of phase 2, $c_t=1$. Then, the selection of $c_{t+1}$, as well as the updating of $\wv$, $k_1$ and $k_2$, depends on $c_{t}$ and $\vv_{t}$, and is described as follows.

\begin{itemize}
    \item[(a)] \textbf{$c_t=1$.} First, we note that since the transmitted symbol is from $w_1$ only, $k_2$ will stay the same. We further divide this case into two subcases: 1) $v_{1,t}=1$. We increase $k_1$ by one, and then compare it with $K$. If it equals $K$, it indicates that update $w_1$ is delivered to user 1. We will then update $w_1$ to the new update generated at next time slot, i.e., $\uv_{t+1}$, and reset $k_1$ to 0.  At same time, if $k_2=K$, we will update $w_2$ to $\uv_{t+1}$, and the system enters phase 1 of the new updating cycle; otherwise, the system stays in phase 2. If $k_1<K$, we keep $w_1,w_2$ unchanged, and let $c_{t+1}=1$. 2) $v_{1,t}=0$. Then, $k_1$, $w_1$, $w_2$ will stay the same. If $v_{2,t}=1$, $c_{t+1}=2$; otherwise, $c_{t+1}=1$.
    \item[(b)] \textbf{$c_t=2$.} For user $i$, if $v_{i,t}=1$, we will increase $k_i$ by one, and then compare with $K$. We consider the following subcases: 1) $v_{1,t}=1$. Similar to the case $c_t=1$, if $k_1=k_2=K$, we update $w_1$ and $w_2$ to $\uv_{t+1}$, and the system enters phase 1 of the new updating cycle; Otherwise, it stays in phase 2 of the current cycle with $c_{t+1}=1$. If $k_1=K$, we update $w_1$ to $\uv_{t+1}$, reset $k_1=0$.  2) $v_{1,t}=0$. We keep $w_1$, $w_2$ the same, and let $c_{t+1}=2$. 
\end{itemize}
\end{itemize}

The procedure is summarized in Algorithm~\ref{algorithm:coding}. 
The adaptive coding scheme has two important features: First, we note that the source transmits three types of symbols: a coded symbol of $w_1$ (which equals $w_2$) in phase 1, an uncoded symbol $w_1(k_1+1)$, or a mixture of $w_1(k_1+1)$ and $w_2$ in phase 2. Since $w_2$ is already decoded by user 1 at the end of phase 1, once user 1 successfully receives any type of such symbols, it accumulates one more novel equation regarding $w_1$. Besides, once it successfully accumulates $K$ equations of $w_1$ and decodes it, the source will switch to a new update immediately. Thus, the adaptive coding scheme works in a greedy fashion for user 1. 

Second, in each updating cycle, user 2 only decodes one update $w_2$. It leverages the diversity of channel conditions to accumulates novel information regarding $w_2$ in phase 2. Specifically, in phase 2, the source would only broadcast a random mixture of $w_1(k_1+1)$ and $w_2$ after user 2 successfully receives $w_1(k_1+1)$ and user 1 has not received it yet. Thus, once user 2 receives such an encoded symbol, it can stripe $w_1(k_1+1)$ away from the mixture, and obtain another novel equation regarding $w_2$. By judiciously selecting the broadcast symbols, we ensure that the information received by user 2 does not involve too many unknown variables, avoiding unnecessary decoding delay.

\begin{algorithm}[t]
\caption{Adaptive Coding and Updating}\label{algorithm:coding}
\begin{algorithmic}[1]
\State{Initialization: $t=1$, $\wv=(\uv_1,\uv_1)$, $k_1=k_2=0$.}
\While{$t$}
\If{$w_1=w_2$}  \Comment{Phase 1}
\State{Send an encoded symbol of $w_1$ and receive $\vv_t$;}
\State{$k_i=\min\{k_i+v_{i,t},K\}$, $i=1,2$;} 
\If{$k_1=K$}
\State{$w_1=\uv_{t+1}$, $k_1=0$;}
\If{$k_2=K$} 
\State{$w_2=\uv_{t+1}$, $k_2=0$;}
\Else \State{$ c_{t+1}=1$;}
\EndIf
\EndIf
\Else  \Comment{Phase 2}
\If{$c_{t}=1$} \Comment{Uncoded symbol}
\State{Send $w_1(k_1+1)$ and receive $\vv_t$;}
\State{$k_1=k_1+v_{i,t}$;} 
\If{$k_1=K$}
\State{$w_1=\uv_{t+1}$, $k_1=0$;}
\If{$k_2=K$} 
\State{$w_2=\uv_{t+1}$, $k_2=0$;}
\EndIf
\EndIf
\If{$v_{1,t}=1$ or $\vv=(0,0)$}
\State{$c_{t+1}=1$;}
\Else
 \State{$c_{t+1}=2$.}
\EndIf
\Else \Comment{Inter-update coding}
\State{Encode $w_1(k_1+1)$ and $w_2$ and transmit.} 
\State{$k_i=\min\{k_i+v_{i,t},K\}$, $i=1,2$;} %
\If{$k_1=K$}
\State{$w_1=\uv_{t+1}$, $k_1=0$;}
\If{$k_2=K$}
\State{$w_2=\uv_{t+1}$, $k_2=0$;}
\EndIf
\EndIf
\If{$v_{1,t}=1$}
\State{$ c_{t+1}=1$;}
\Else
\State{$ c_{t+1}=2$;}
\EndIf
\EndIf
\EndIf
\State{$t=t+1$;}
\EndWhile
\end{algorithmic}
\end{algorithm}

\section{Analysis of the AoI at both Users}\label{sec:coding_ana}
First, we note that under the adaptive coding scheme, the source always broadcasts a new equation involving the update user 1 demands in each time slot. Therefore, the AoI of user 1 evolves exactly the same as under the greedy I scheme. Thus, the long-term average AoI of user 1 remains unchanged, which is equal to $\Delta_1=\frac{K}{p_1}\left(\frac{3}{2}+\frac{1-p}{K}\right)$.

Next, we will analyze the AoI of user $2$ under the adaptive coding scheme. We note that the resulted updating cycles form a renewal process, where each renewal interval begins when both users demand the same updates (i.e., $w_1=w_2$), and ends when user 1 successfully decodes an update after use 2 decodes $w_2$. As shown in Fig.~\ref{fig:coding}, we further decompose each renewal interval into three different stages: phase 1, phase 2a, which begins when the system enters phase 2 and ends when user 2 decodes $w_2$, and phase 2b, the duration user 1 takes to complete the current update $w_1$ after user 2 decodes $w_2$. We denote the lengths of those stages as $T_1$, $T_2$ and $T_3$, respectively. In the following, we will analyze each of them individually, and then obtain an upper bound on $\Delta_2$.

\subsection{Analysis of $T_1$}
Since $T_1$ follows a negative binomial distribution with parameter $K,p_1$, we have 
\begin{align}
&\Eb[T_1]=\frac{K}{p_1},\quad \Eb[T_1^2]=\frac{K^2}{p_1^2}+\frac{(1-p_1)K}{p_1}.\label{eqn:T_1^2}
\end{align}

\subsection{Analysis of $T_2$}
 Let $\bar{k}_2$ be the number of symbols delivered to user 2 during phase 1. Then, $\bar{k}_2=\min\left\{\sum_{t=1}^{T_1} v_{2,t}, K\right\}.$
 The system will enter phase 2 if $\bar{k}_2<K$.
 
\begin{figure}
	\centering
	\includegraphics[width=3.5in]{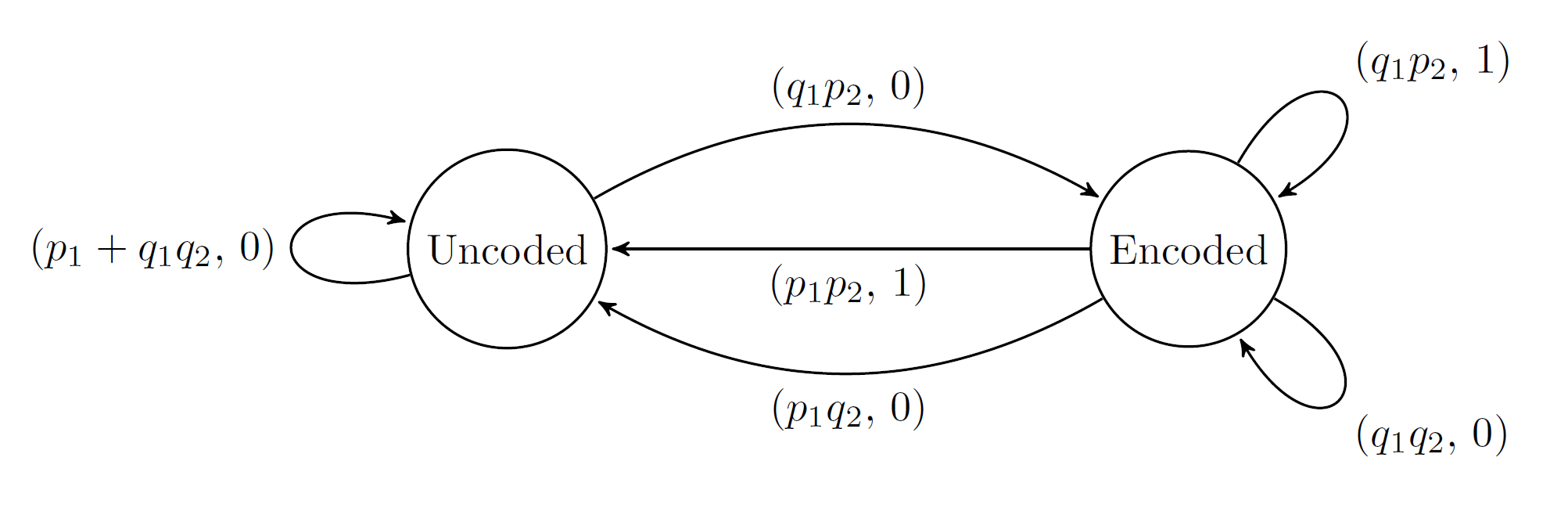}
	\captionof{figure}{Associated Markov chain in phase 2.}
	\label{fig:markov}
	\vspace{-0.15in}
\end{figure}

 In order to analyze $T_2$, we first introduce a Markov chain associated with the coding and updating process in phase 2, as shown in Fig.~\ref{fig:markov}. The Markov chain has two states named ``uncoded" and ``encoded", corresponding to the encoding decisions $c_t=1$ and $c_t=2$, respectively. The evolution of the Markov chain depends on the transmission results $\vv_t$, similar to the coding scheme in phase 2.
 
 In order to track the number of equations user 2 receives regarding $w_2$ in phase 2, we associate a reward with each transition of the Markov chain. The reward denotes the increment of $k_2$ after each transmission. As depicted in Section~\ref{sec:coding}, $k_2$ will increase by one at the end of time slot $t$ if $c_t=2$ and $v_{2,t}=1$. This is because $x_t$ is a linear combination of $w_1(k_1+1)$ and $w_2$, and $w_1(k_1+1)$ has been successfully received by user 2 previously. Thus, after the successful transmission of $x_t$, user 2 can strip $w_1(k_1+1)$ away from $x_t$, and obtain a new linear equation about $w_2$. The transition probabilities and the associated rewards are shown in Fig.~\ref{fig:markov}. 

Phase 2a) begins at state ``uncoded", and ends when user 2 accumulates $K$ equations regarding $w_2$ and successfully decodes it. Although the Markovian structure admits a closed-form stationary distribution, the non-asymptotic analysis of $T_2$ is not straightforward. To make it tractable, we will consider a renewal reward process embedded in the Markov structure, and leverage tools such as stopping time theory to analyze it.

\begin{figure}[t]
	\centering
	\includegraphics[width=2.4in]{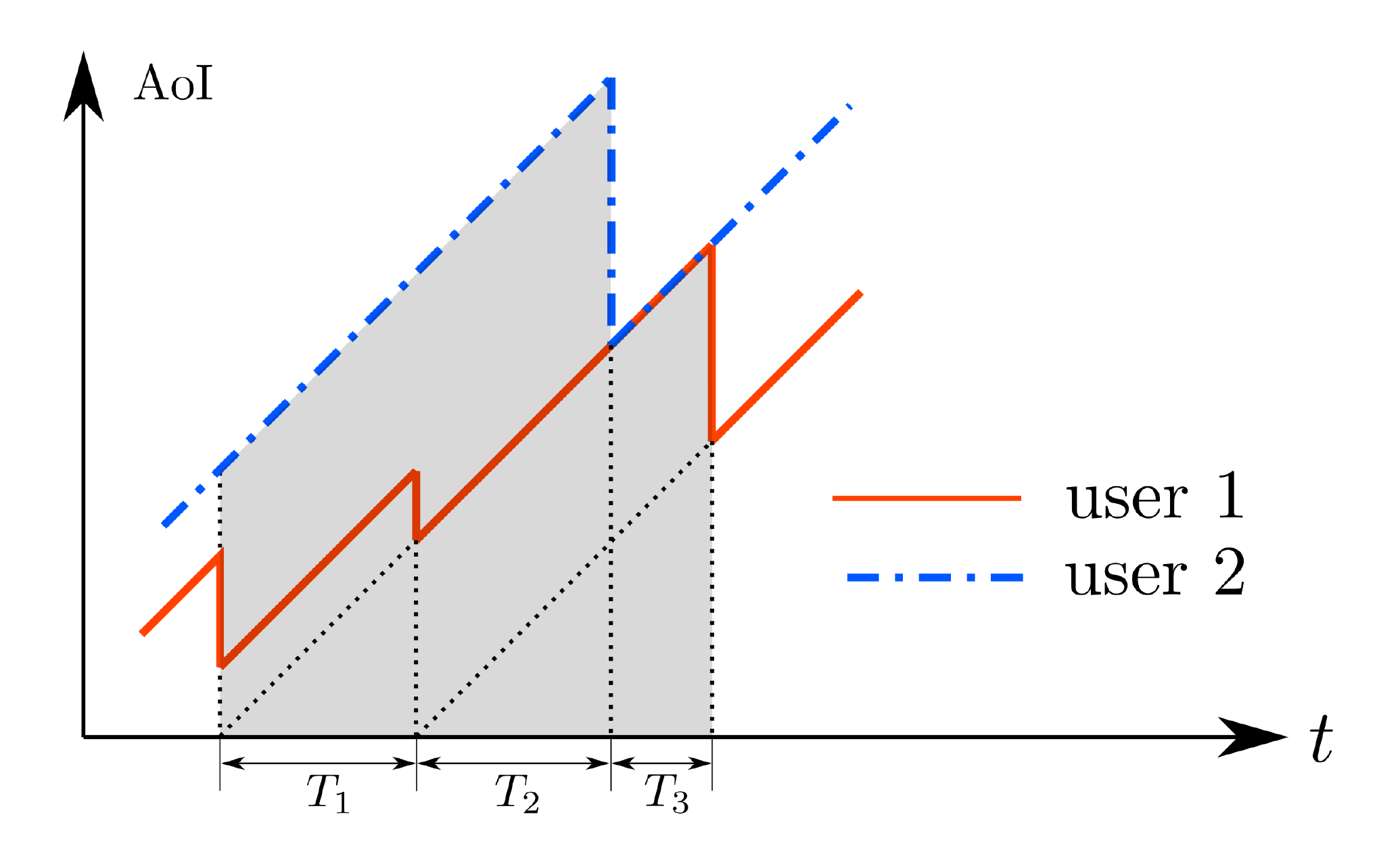}
	\captionof{figure}{An illustration of the updating cycle.}
	\label{fig:coding}
	\vspace{-0.15in}
\end{figure}

Define a renewal process where each renewal interval corresponds to the duration between two consecutive visits to state ``uncoded" under the Markov chain. Let $\{Z_i\}_i$ be the lengths of the renewal intervals, and $\{W_i\}_i$ be the total rewards (i.e., total increments of $k_2$) over individual renewal intervals. Then, we have the following observations.
\begin{Proposition}
Let $V_i$ be a geometric random variable with parameter $p_1$, and $R_t$ be i.i.d. Bernoulli random variables with parameter $p_2$. Then, $(Z_i,W_i)$ are i.i.d. random pairs with
\begin{align*}
(Z_i,W_i)&=\left\{ \begin{array}{ll}
(1,0),& \mbox{w.p. } \quad p_1+q_1q_2 ,\\
(1+V_i,\sum_{t=1}^{V_i} R_t) & \mbox{w.p. } \quad 1-(p_1+q_1q_2),
\end{array}\right.
\end{align*}
where $q_i:=1-p_i$, $i=1,2$.
\end{Proposition}
\begin{Proof}
    Under the Markov chain, $Z_i=1$ if $v_{1,t}=1$ or $\vv_t=(0,0)$, which happens with probability $p_1+q_1q_2$. Otherwise, the system enters state ``encoded" and stays there until user 1 successfully receives an encoded symbol (i.e., $v_{1,t}=1$). The total duration the system stays in state ``encoded", denoted as $V_i$, is thus a geometric random variable with parameter $p_1$. The reward obtained over $Z_i$ is thus equals to the total number of successful transmissions when the system stays in the state  ``encoded", which is the summation of $V_i$ i.i.d. Bernoulli random variables with probability $p_2$.
\end{Proof}

\begin{Proposition}\label{prop:T_2} $T_2$ is upper bounded by $\sum_{i=1}^{N_1} Z_i$, where $N_1$ is a stopping time determined by
$N_1 =\min \left\{n\middle|  \sum_{i=1}^n W_i \geq K \right\}$.
\end{Proposition}
\begin{Proof}
    Under the original coding and updating process, when the system enters phase 2a, we have $0\leq \bar{k}_2<K$. Phase 2a ends as soon as the cumulative number of equations received by user 2 regarding $w_2$ reaches $K$. Thus, $T_2$ will be upper bounded by $\sum_{i=1}^{N_1} Z_i$ under each sample path. Since $N_1$ only depends on observed $W_i$s, it is a stopping time.
\end{Proof}

We point out that $T_2$ essentially depends on phase 1 through $\bar{k}_2$. The upper bound of $T_2$ removes such dependency by relaxing $\bar{k}_2$ to zero.

\begin{Lemma}\label{lemma:Z} The first and second moments of $Z_i$ equal
\begin{align*} 
\Eb[Z_i]&= 1+\frac{q_1p_2}{p_1}, \quad\Eb[Z_i^2]=1-p_2+\frac{4p_2}{p_1^2}.
\end{align*}
\end{Lemma}
The proof of Lemma~\ref{lemma:Z} is omitted due to space limitation.

Next, we will derive proper bounds for $\Eb[T_2]$ and $\Eb[T_2^2]$.
In order to simplify the analysis, we consider \emph{capped} rewards $\bar{W}_i$ instead of $W_i$. Specifically, we let $\bar{W}_i=\min\{W_i,1\}.$ We also define another stopping time $\bar{N}_{1}$ as

We have the following observation.
\begin{Lemma}\label{lemma:N}
The capped reward process $\{\bar{W}_i\}$ is an i.i.d. Bernoulli process with parameter $r:=p_1+q_1q_2+\frac{p_1p_2{q_1q_2}}{1-q_1q_2}$. Besides,  \begin{align*}
\Eb[\bar{N}_{1}]&=\frac{K}{1-r}, \quad \Eb[\bar{N}_{1}^2]=\frac{K(K+r)}{(1-r)^2}.
\end{align*}
\end{Lemma}
\begin{Proof}
   Based on the definitions of $W_i$ and $\bar{W}_i$, we have
\begin{align*}
&\Pb[\bar{W}_i=0]=\Pb[{W}_i=0]\\
&=p_1+q_1q_2+q_1p_2\sum_{l=0}^\infty (q_1q_2)^l p_1q_2:=r.
\end{align*}
Since $\bar{N}_1$ is essentially a negative binomial random variable with parameters $K$ and $r$, its first and second moments can thus be easily derived.
\end{Proof}

Before we proceed to bound the first and second moments of $\sum_{i=1}^{\bar{N}_1}Z_i$, we introduce the following lemma.

\begin{Lemma}[Sharp Moment Inequality from \cite{Pawel:Math}]
Let $\{X_t\}$ be a sequence of independent non-negative random variables, $\tau$ be a stopping time, and $\tau'$ be a copy of $\tau$ independent with $\{X_t\}$. Then,
\begin{align*}
    \Eb\bigg(\sum_{i=1}^{\tau}Z_i\bigg)^p \leq 2^{p-1}\Eb\bigg(\sum_{i=1}^{\tau'}Z_i\bigg)^p,\quad 1\leq p<\infty.
\end{align*}\label{lemma:sharp}
\end{Lemma}

Lemma~\ref{lemma:sharp} enables us to decouple the dependency between $\bar{N_1}$ and $\{Z_i\}$ to obtain the corresponding upper bounds as follows.

\begin{Lemma}\label{lemma:T_2_bound}
Based on the definitions of $Z_i$, $\bar{N}_1$, we have
\begin{align*}
\Eb\left[\sum_{i=1}^{\bar{N}_1}Z_i\right]&=\frac{K}{1-r}\left(1+\frac{q_1p_2}{p_1}\right),\\
\Eb\bigg(\sum_{i=1}^{\bar{N}_1}Z_i\bigg)^2&\leq 
\frac{2K(K+2r-1)}{(1-r)^2}\left(1+\frac{q_1p_2}{p_1}\right)^2 \\ &\quad+\frac{2K}{1-r}\left(1-p_2+\frac{4p_2}{p_1^2}\right).
\end{align*}
\end{Lemma}
The first equality in Lemma~\ref{lemma:T_2_bound} can be proved based on Wald's identity. The bound on the second moment is based on Lemma~\ref{lemma:sharp} by setting $p=2$, and the results from Lemma~\ref{lemma:Z} and Lemma~\ref{lemma:N}.

\textbf{Remark:} Since $\bar{W}_i$ is a capped version of $W_i$, $\bar{N}_1\geq N_1$ under every sample path. Thus, the results in Lemma~\ref{lemma:T_2_bound} serve as upper bounds for $\Eb\left[\sum_{i=1}^{{N}_1}Z_i\right]$ and $\Eb\big(\sum_{i=1}^{\bar{N}_1}Z_i\big)^2$, which also upper bound $\Eb[T_1]$ and $\Eb[T_1^2]$ according to Proposition~\ref{prop:T_2}. Therefore, we have $\Eb[T_1]=O(K)$ and $\Eb[T_1^2]=O(K^2)$.

\subsection{Analysis of $T_3$}
Since $T_3$ is the remaining time that user 1 takes to complete current update after user 2 decodes $w_2$, the remaining symbol user 1 demand is upper bounded by $K$. Therefore, $T_3$ is bounded by the summation of $K$ i.i.d. geometric random variables with parameter $p_1$. $\Eb[T_3]$ and $\Eb[T_3^2]$ can thus be bounded in the same way as $T_1$.

\subsection{Bound $\Delta_2$}
As illustrated in Fig.~\ref{fig:coding}, under the adaptive coding scheme, we have
\begin{align*}
\Delta_2
&\leq \Eb[{T}_1+{T}_2+T_3]+\frac{\Eb[(T_1+T_2+T_3)^2]}{2\Eb[T_1+T_2+T_3]}.
\end{align*}
Combining the results on $T_1$, $T_2$ and $T_3$, we obtain the following theorem.

\begin{Theorem}\label{thm:coding}
Under the adaptive coding scheme, we have $\Delta_1=\frac{K}{p_1}\left(\frac{3}{2}+\frac{1-p}{K}\right)$, $\Delta_2=O(K)$.
\end{Theorem}
Theorem~\ref{thm:coding} indicates that, compared with the greedy scheme in Sec.~\ref{sec:greedy}, the adaptive coding scheme improves the AoI at user 2 from $\Omega((1/{q})^{K} K)$ to $O(K)$ without affecting the AoI at user 1.

\section{Numerical Results}\label{sec:simu}
In this section, we evaluate the proposed coding schemes through simulation. We plot the sample average of AoI over $50$ sample paths.

\begin{figure}
	\centering
	\vspace{-0.1in}
	\includegraphics[width=2.9in]{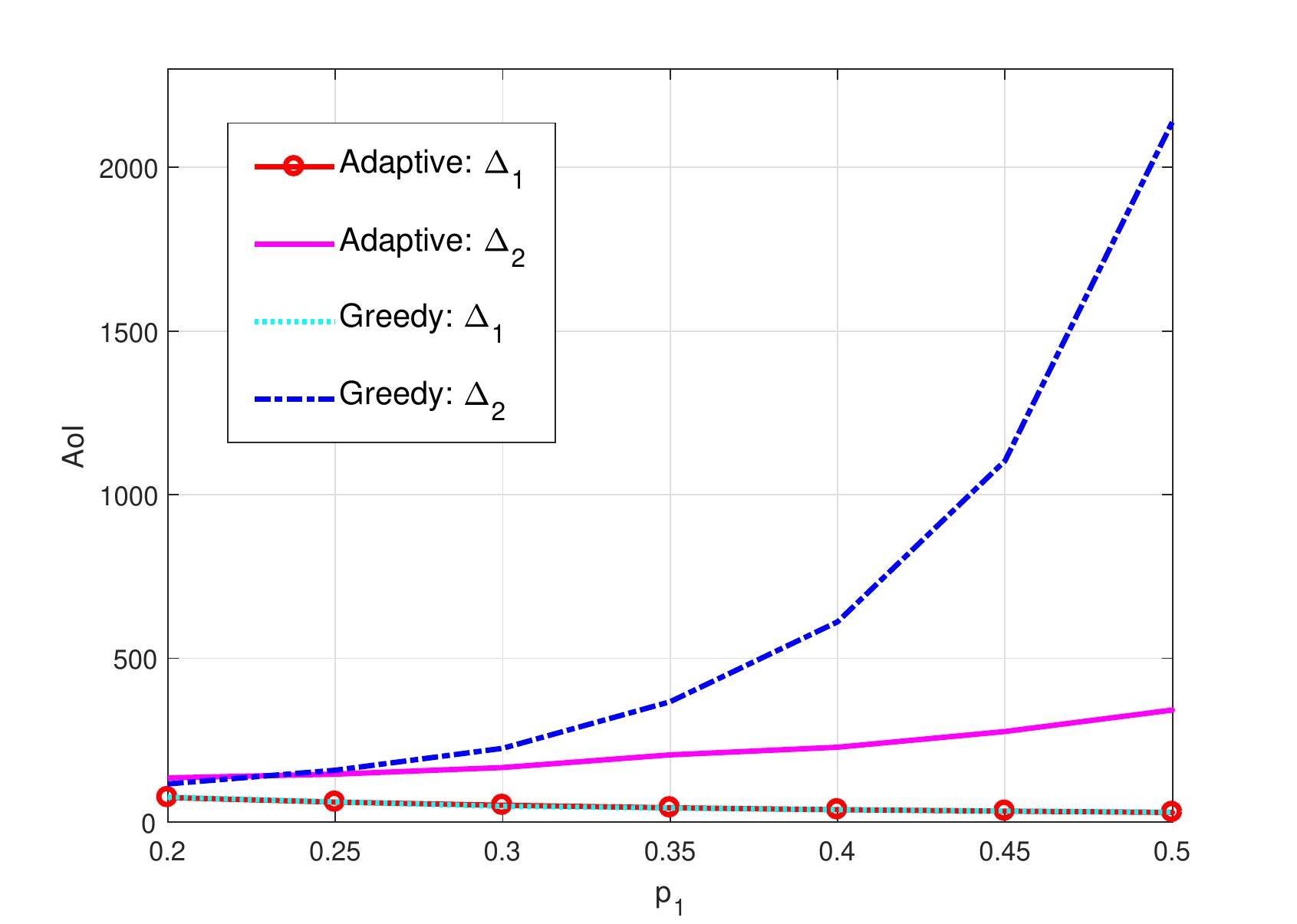}
		\vspace{-0.1in}
	\captionof{figure}{AoI as a function of $p_1$ when $p_2=0.2, K=10$.}
	\vspace{-0.2in}
	\label{fig:simu-1}
\end{figure}

First, we fix $K=10$, $p_2=0.2$ and vary $p_1\in(0.2, 0.5]$. The corresponding AoI under the greedy scheme and the adaptive coding scheme is depicted in Fig.~\ref{fig:simu-1}. We note that the AoI at user 1 under the adaptive coding scheme matches that under the greedy scheme. Besides, the average AoI at user $2$ increases in $p_1$ in a super-linear fashion under the greedy scheme, while it only increases approximately linearly under the adaptive coding scheme. This indicates that the coding gain is more prominent when the channel qualities of both users differ more.

\begin{figure}
	\centering
	\includegraphics[width=2.9in]{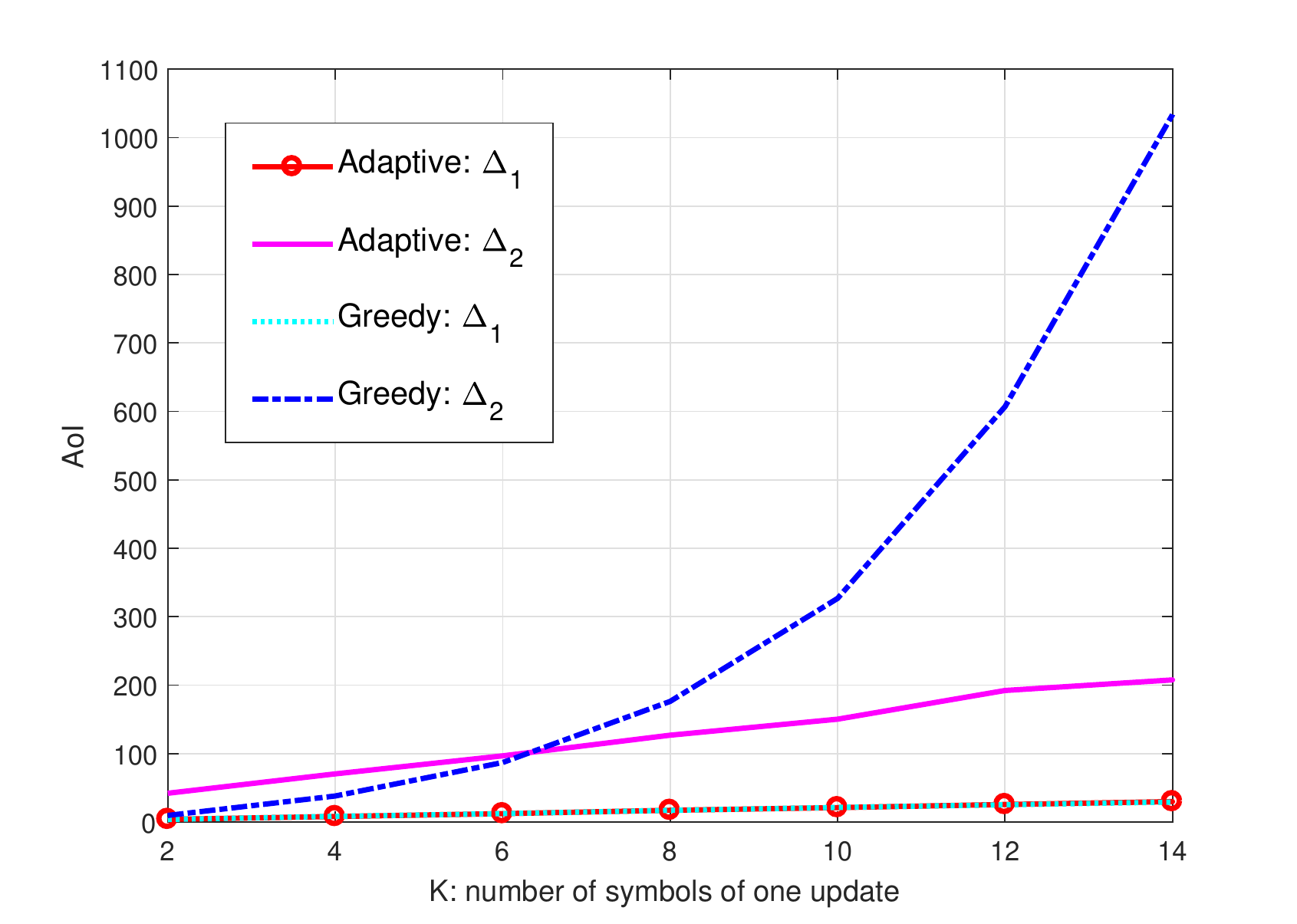}
	\captionof{figure}{AoI as a function of $K$ when $p_1=0.7, p_2=0.4$.}
		\vspace{-0.2in}
	\label{fig:simu-3}
\end{figure}

Next, we fix $p_1=0.7$ and $p_2=0.4$ and evaluate the sample average AoI with different size of the update $K$. We observe similar results in Fig.~\ref{fig:simu-3}. We note that the AoI at user 2 increases super-linearly in $K$ under the greedy scheme, and only scales linearly in $K$ under the adaptive coding scheme, which corroborates the theoretical results in Theorem~\ref{thm:greedy} and Theorem~\ref{thm:coding}.

\section{Discussion and Extension}
In this work, we demonstrated the benefits of adaptive coding in a two-user broadcast symbol erasure channel with feedback. The future directions include extending the analysis to other coding schemes and general $M$-user broadcast channels.

\if{0}
\begin{center}
\begin{tabular}{|c|c|c|}
\hline 
Scheme & $\Delta_1$ & $\Delta_2$  \\ \hline
Greedy scheme I & $\frac{K}{p_1}\left(\frac{3}{2}+\frac{1-p_1}{K}\right)$ & $\Omega\left(\frac{K}{\bar{q}^K}\right)$    \\ \hline
Greedy scheme II & $\Omega(K)$ & $\frac{K}{p_2}\left(\frac{3}{2}+\frac{1-p_2}{K}\right)$    \\ \hline
Greedy scheme III & $\Omega(K)$ & $\Omega\left(\frac{K}{\bar{q}^K}\right)$    \\ \hline
Inter-Updating Coding & $\frac{K}{p_1}\left(\frac{3}{2}+\frac{1-p_1}{K}\right)$ & O(K)    \\ \hline
\end{tabular}
\end{center}
\fi

\bibliographystyle{IEEEtran}
\bibliography{AgeInfo,ener_harv}

\end{document}